\documentclass{article}
\usepackage{amssymb}
\usepackage{amsmath}
\usepackage[english]{babel}
\newtheorem{theorem}{Theorem}

\newtheorem{lemma}[theorem]{Lemma}
\newtheorem{corollary}[theorem]{Corollary}
\newtheorem{remark}[theorem]{Remark}
\begin{document}
\title{Semiclassical spectral estimates for Schr\"odinger operators at a
critical energy level. Case of a degenerate minimum of the
potential}
\author{Brice Camus\\Ruhr-Universit\"at Bochum. Fakult\"at f\"ur
Mathematik,\\
Universit\"atsstr. 150, D-44780 Bochum, Deutschland.\\Email :
brice.camus@univ-reims.fr} %
\date{Received: date / Accepted: date}
\maketitle
\begin{abstract}
We study the semi-classical trace formula at a critical energy
level for a Schr\"odinger operator on $\mathbb{R}^{n}$. We assume
here that the potential has a totally degenerate critical point
associated to a local minimum. The main result, which computes the
contribution of this equilibrium, is valid for all time in a
compact and establishes the existence of a total asymptotic
expansion whose top order coefficient depends only on the germ of
the potential at the critical point.
\end{abstract}
\section{Introduction.}
Let us consider $P_{h}$ a self-adjoint $h$-pseudo-differential
operator, or more generally $h$-admissible (see \cite{[Rob]}),
acting on a dense subset of $L^2(\mathbb{R}^n)$. A classical
problem is to study the asymptotic behavior, as $h$ tends to 0, of
the spectral function :
\begin{equation}
\gamma (E,h,\varphi )=\sum\limits_{|\lambda _{j}(h)-E|\leq
\varepsilon }\varphi (\frac{\lambda _{j}(h)-E}{h}),  \label{Def
trace}
\end{equation}
where the $\lambda _{j}(h)$ are the eigenvalues of $P_{h}$, $E$ is
an energy level of the principal symbol of $P_h$ and $\varphi$ a
function. Here we suppose that the spectrum is discrete in
$[E-\varepsilon ,E+\varepsilon ]$, a sufficient condition for this
is given below. If $p_0$ is the principal symbol of $P_{h}$ we
recall that an energy $E$ is regular when $\nabla p_0(x,\xi )\neq
0$ on the energy surface :
\begin{equation}
\Sigma _{E}=\{(x,\xi )\in T^{\star}\mathbb{R}^{n}\text{ }/\text{
}p_0(x,\xi )=E\},
\end{equation}
and critical when it is not regular. A well established result is
the existence of a link between the asymptotics of (\ref{Def
trace}), as $h$ tends to 0, and the closed trajectories of the
Hamiltonian flow of $p_0$ on the energy surface $\Sigma_{E}$. This
duality between spectrum and periodic orbits exists also in a
non-semiclassical context such as in the Selberg trace formula or
for the trace of the wave operator on compact manifolds. In the
semi-classical setting this kind of correspondence was initially
pointed out in the physic literature : Gutzwiller \cite{GUT},
Balian\&Bloch \cite{BB}. From a mathematical point of view, and
when $E$ is a regular energy, a non-exhaustive list of references
is Brummelhuis\&Uribe \cite{BU}, Petkov\&Popov \cite{P-P},
Paul\&Uribe \cite{PU} and more recently Combescure et al.
\cite{CRR} with a different approach based on coherent states.

When one drops the assumption that $E$ be a regular value, and
this will be the case here, the behavior of (\ref{Def trace})
depends on the nature of the singularities of $p$ on $\Sigma_{E}$.
This problem is too complicated in general position and some extra
hypotheses on $p$ are required. The case of a non-degenerate
critical energy, i.e. such that the critical-set
$\mathbb{\frak{C}}(p_0) =\{(x,\xi )\in T^{\ast
}\mathbb{R}^{n}\text{ }/\text{ }dp_0(x,\xi )=0\}$ is a compact
$C^{\infty }$ manifold with a Hessian $d^{2}p_0$ transversely
non-degenerate along this manifold, has been investigated first by
Brummelhuis et al. in \cite{BPU}. They treated this question for
quite general operators but it was assumed that 0 was the only
period of the linearized flow in $\rm{supp}(\hat{\varphi})$. The
case of the non-zero periods of the linearized flow was obtained
by Khuat-Duy \cite{KhD1} with $\rm{supp}(\hat{\varphi})$ compact,
but for Schr\"{o}dinger operators with non-degenerate potentials.
Our contribution was to generalize his result for some more
general operators, but under some geometrical assumptions on the
flow (see \cite{Cam}). Finally, in \cite{Cam1,Cam2} the case of
degenerate critical points for $h$-pseudo-differential operators
was obtained respectively for elliptic and real-principal type
singularities but with a very restrictive assumption on the
dimension for the later. In this work we treat the case of
Schr\"odinger operators near a degenerate minimum of the
potential.

After a reformulation, the spectral function (\ref{Def trace}) can
be expressed in terms of oscillatory integrals whose phases are
related to the Hamiltonian flow $\Phi_{t}=\mathrm{exp}(tH_{p_0})$.
Precisely, the asymptotic behavior as $h$ tends to 0 of these
oscillatory integral is related to the closed orbits of this flow.
When $(x_0,\xi_0)$ is a critical point of $p_0$, and hence an
equilibrium of the flow, it is well known that the relation :
\begin{equation}\label{cond.IO.singular}
\mathfrak{F}_t=\mathrm{Ker}(d_{x,\xi}\Phi_{t}(x_0,\xi_0)-\mathrm{Id})\neq
\{ 0\},
\end{equation}
leads to the study of degenerate oscillatory integrals. In this
work we consider :
\begin{equation}
P_h=-h^2\Delta+V(x),
\end{equation}
but the main result can be applied to an $h$-admissible operator,
of principal symbol $\xi^2+V(x)$. In particular, we investigate
the case of a smooth potential $V$ with a single and degenerate
critical point $x_0$ attached to a local minimum. An immediate
consequence is that the symbol admits a unique critical point
$z_0=(x_0,0)$ and that the linearized flow at this point is given
by the flow of the free Laplacian. Also, Eq.
(\ref{cond.IO.singular}) is automatically satisfied with :
\begin{equation*}
\mathfrak{F}_t=\{ (x,\xi)\in T_{z_0}T^{*}\mathbb{R}^n \text{ /
}\xi=0\}, \text{ }t\neq 0,
\end{equation*}
and $\mathfrak{F}_0=T_{z_0}T^{*}\mathbb{R}^n\simeq
\mathbb{R}^{2n}$. Hence, the stationary phase method cannot be
applied, same in a version with parameter, in a neighborhood of
$t=0$.

The proof is based on the BKW ansatz and on the existence of
suitable local normal forms near the equilibrium for our phase
functions with a generalization of the stationary phase formula
for these normal forms. This generalization holds only in the case
of a local minimum of the potential and cannot be applied, for
example, to the case of a local maximum of the potential.
\section{Hypotheses and main result.}
Let be $p(x,\xi)=\xi^2 +V(x)$ where the potential $V$ is smooth on
$\mathbb{R}^n$ and real valued. To this symbol is attached the
$h$-differential operator $P_h=-h^2\Delta+V(x)$ on
$C_0^{\infty}(\mathbb{R}^n)\subset L^2(\mathbb{R}^n)$. Clearly,
$P_h$ is essentially autoadjoint if $V$ is positive. To have a
well defined spectral problem we use :\medskip\\
$(H_{1})$ \textit{There exists } $\varepsilon_{0}>0$ \textit{ such that }%
$p^{-1}([E-\varepsilon_{0},E+\varepsilon_{0}])$\textit{\ is
compact.}\medskip\\
For example $(H_{1})$ is certainly satisfied if $V$ tends to
infinity at infinity. By Theorem 3.13 of \cite{[Rob]} the spectrum
$\sigma (P_{h})\cap [E-\varepsilon ,E+\varepsilon ]$ is discrete
and consists in a sequence $\lambda _{1}(h)\leq \lambda
_{2}(h)\leq ...\leq \lambda _{j}(h)$ of eigenvalues of finite
multiplicities if $\varepsilon$ and $h$ are small enough. In this
setting, and with $E_c$ critical, we want to study the asymptotics
of the spectral function :
\begin{equation}
\gamma (E_{c},h,\varphi)=\sum\limits_{\lambda _{j}(h)\in
[E_{c}-\varepsilon ,E_{c}+\varepsilon ]}\varphi (\frac{\lambda
_{j}(h)-E_{c}}{h}). \label{Objet trace}
\end{equation}
To avoid any problem of convergence we impose the
classical condition : \medskip\\
$(H_{2})$\textit{ We have }$\hat{\varphi}\in C_{0}^{\infty
}(\mathbb{R}).$\ \medskip\\
To simplify, we will write $z=(x,\xi)$ for any point of the phase
space. Let be $\Sigma _{E_{c}}=p^{-1}(\{E_{c}\})$, the
singularities we consider are :\medskip\\
$(H_{3})$\textit{ On }$\Sigma _{E_{c}}$ \textit{the symbol}
$p$\textit{ has a unique critical point }$z_{0}=(x_{0},0).$
\textit{This critical point is degenerate, of finite order,
associated to a local minimum of the potential
$V$ and the first non-zero homogeneous term of the germ of $V$ in $x_0$ is definite positive.}\medskip\\
By degenerate we mean that the Hessian of $V$ in $x_0$ is zero.
Hence, $(H_{3})$ insures that the germ of $V$ can be written as :
\begin{equation}\label{form pot}
V(x)=E_c+\sum\limits_{j=k}^{N}
V_j(x)+\mathcal{O}(||x-x_0||^{N+1}),
\end{equation}
where each $V_j$ is homogeneous of degree $j$ w.r.t. $(x-x_0)$,
$k\geq 4$ is even and $V_k$ is definite positive.
\begin{remark}
\rm{Since all previous derivatives are 0 in $x_0$, $V_k$ is
invariantly defined w.r.t. the choice of local coordinates on
$\mathbb{R}^n$ near $x_0$. Also, since $\xi^2>0$ for $\xi\neq 0$,
elementary considerations show that $(x_0,0)$ is isolated on
$\Sigma_{E_c}$.}$\hfill{\square}$
\end{remark}
Let be $\Theta$ a cutoff function near the critical energy $E_c$
and $\psi \in C_{0}^{\infty }(T^{\ast }\mathbb{R}^{n})$
microlocally supported near $z_{0}$. To understand the
contribution of the equilibrium it suffices to study the
microlocal problem :
\begin{equation}
\gamma _{z_{0}}(E_{c},h,\varphi)=\frac{1}{2\pi }\mathrm{Tr}\int\limits_{\mathbb{R}}e^{i%
\frac{tE_{c}}{h}}\hat{\varphi}(t)\psi ^{w}(x,hD_{x})\mathrm{exp}(-\frac{it}{h}%
P_{h})\Theta (P_{h})dt.
\end{equation}
See section 3 below for the introduction of $\Theta (P_{h})$ and
section 4 for $\psi^{w}(x,hD_{x})$. The main result of the present
work is :
\begin{theorem}\label{Main}
Under hypotheses $(H_{1})$ to $(H_{3})$ we have :
\begin{equation*}
\gamma _{z_{0}}(E_{c},h,\varphi)\sim
h^{-n+\frac{n}{2}+\frac{n}{k}} \sum\limits_{j,l\in\mathbb{N}^2}
h^{\frac{j}{2}+\frac{l}{k}}\Lambda _{j,l}(\varphi ),
\end{equation*}
where the $\Lambda _{j,l}$ are some computable distributions. The
leading coefficient is :
\begin{equation}
\Lambda _{0,0}(\varphi )
=\frac{\mathrm{S}(\mathbb{S}^{n-1})}{(2\pi)^n}
\int\limits_{\mathbb{S}^{n-1}} |V_k(\eta)|^{-\frac{n}{k}} d\eta
\int\limits_{\mathbb{R}_{+} \times \mathbb{R}_{+}} \varphi(u^2
+v^k) u^{n-1} v^{n-1} dudv,
\end{equation}
where $\mathrm{S}(\mathbb{S}^{n-1})$ is the surface of the
unit-sphere of $\mathbb{R}^n$.
\end{theorem}
The leading term of Theorem \ref{Main} differs from the regular
case. For $E$ non-critical and $\mathrm{supp}(\hat{\varphi})$
small enough near the origin a classical estimates, see e.g.
\cite{PU}, is :
\begin{equation}
\gamma(E,h,\varphi)= (2\pi h)^{1-n} (\hat{\varphi}(0)
\mathrm{LVol}(\Sigma_E) +\mathcal{O}(h)),
\end{equation}
where $\mathrm{LVol}(\Sigma_E)$ is the Liouville measure of the
compact energy surface $\Sigma_E$. Hence, for $n>1$ the main term
of Theorem \ref{Main} is smaller since
$\frac{n}{2}+\frac{n}{k}>1$.\medskip\\
This result can be interpreted if we consider high energies
estimates :
\begin{equation} \label{toy}
\mathrm{Tr}\text{ } \varphi(\frac{-\Delta +V_k(x)}{h}).
\end{equation}
By the Weyl-law, we know that the trace exists if $\varphi$
decreases fast enough at infinity. This trace can be computed if
we scale $h$ out via :
\begin{gather*}
\int\limits_{\mathbb{R}^{2n}} \varphi(\frac{\xi^2
+V_k(x)}{h})dxd\xi=
h^{\frac{n}{2}+\frac{n}{k}}\int\limits_{\mathbb{R}^{2n}}
\varphi(\xi^2 +V_k(x))dxd\xi\\
=\mathrm{S}(\mathbb{S}^{n-1})
h^{\frac{n}{2}+\frac{n}{k}}\int\limits_{\mathbb{R}_{+}^2}
\varphi(r^2 +q^k) r^{n-1}q^{n-1} drdq
\int\limits_{\mathbb{S}^{n-1}} |V_k(\eta)|^{-\frac{n}{k}} d\eta.
\end{gather*}
Hence, the main contribution in the expansion of
$\gamma_{z_0}(E_c,h,\varphi)$ is just $(2\pi h)^{-n}$ times the
high energies estimate of Eq. (\ref{toy}). But Theorem \ref{Main}
shows more since we have the existence of a complete asymptotic
expansion for $\gamma_{z_0}(E_c,h,\varphi)$. Finally, the
contribution studied in Theorem \ref{Main} is exactly
$\gamma(E_c,h,\varphi)$ if the critical point is a strict minimum
of the potential since in this case $\Sigma_{E_c}=\{z_0\}$.
\section{Oscillatory representation.}
Let be $\varphi \in \mathcal{S}(\mathbb{R})$ with
$\hat{\varphi}\in C_{0}^{\infty }(\mathbb{R})$, we recall that :
\begin{equation*} \gamma (E_{c},h,\varphi)
=\sum\limits_{\lambda _{j}(h)\in I_{\varepsilon }}\varphi
(\frac{\lambda _{j}(h)-E_{c}}{h}), \text{ } I_{\varepsilon }
=[E_{c}-\varepsilon ,E_{c}+\varepsilon ],
\end{equation*}
with $0<\varepsilon<\varepsilon_0$ and
$p^{-1}(I_{\varepsilon_{0}})$ compact in $T^{\ast
}\mathbb{R}^{n}$. We localize near the critical energy $E_{c}$
with a cut-off function $\Theta \in C_{0}^{\infty
}(]E_{c}-\varepsilon ,E_{c}+\varepsilon \lbrack )$, such that
$\Theta =1$ near $E_{c}$ and $0\leq \Theta \leq 1$. The associated
decomposition is :
\begin{equation*}
\gamma (E_{c},h,\varphi)=\gamma _{1}(E_{c},h,\varphi)+\gamma
_{2}(E_{c},h,\varphi),
\end{equation*}
with :
\begin{equation}
\gamma _{1}(E_{c},h,\varphi)=\sum\limits_{\lambda _{j}(h)\in
I_{\varepsilon }}(1-\Theta )(\lambda _{j}(h))\varphi
(\frac{\lambda _{j}(h)-E_{c}}{h}),
\end{equation}
\begin{equation}
\gamma _{2}(E_{c},h,\varphi)=\sum\limits_{\lambda _{j}(h)\in
I_{\varepsilon }}\Theta (\lambda _{j}(h))\varphi (\frac{\lambda
_{j}(h)-E_{c}}{h}).
\end{equation}
As concerns $\gamma _{1}(E_{c},h,\varphi)$ a well known result,
see e.g.\cite{Cam1}, is :
\begin{lemma}
$\gamma _{1}(E_{c},h,\varphi)=\mathcal{O}(h^{\infty })$, as
$h\rightarrow 0$.\label{S1(h)=Tr}
\end{lemma}
Consequently, for the study of $\gamma (E_{c},h,\varphi)$ modulo
$\mathcal{O}(h^{\infty })$, we have only to consider $\gamma
_{2}(E_{c},h,\varphi)$. By inversion of the Fourier transform we
have :
\begin{equation*}
\Theta (P_{h})\varphi (\frac{P_{h}-E_{c}}{h})=\frac{1}{2\pi}\int\limits_{%
\mathbb{R}}e^{i\frac{tE_{c}}{h}}\hat{\varphi}(t)\mathrm{exp}(-\frac{it}{h}%
P_{h})\Theta (P_{h})dt.
\end{equation*}
The trace of the left hand-side is exactly $\gamma
_{2}(E_{c},h,\varphi)$ and Lemma \ref{S1(h)=Tr} gives :
\begin{equation}\label{Trace S2(h)}
\gamma (E_{c},h,\varphi)=\frac{1}{2\pi }\mathrm{Tr}\int\limits_{\mathbb{R}}e^{i%
\frac{tE_{c}}{h}}\hat{\varphi}(t)\mathrm{exp}(-\frac{it}{h}P_{h})\Theta
(P_{h})dt+\mathcal{O}(h^{\infty }).
\end{equation}
Let be $U_{h}(t)=\mathrm{exp}(-\frac{it}{h}P_{h})$, the evolution
operator. For each integer $N$ we can approximate $U_{h}(t)\Theta
(P_{h})$ by a Fourier integral-operator depending on the parameter
$h$. If $\Lambda$ is the Lagrangian manifold associated to the
flow of $p$ :
\begin{equation*}
\Lambda =\{(t,\tau ,x,\xi ,y,\eta )\in T^{\ast }\mathbb{R}\times T^{\ast }%
\mathbb{R}^{n}\times T^{\ast }\mathbb{R}^{n}:\tau =p(x,\xi
),\text{ }(x,\xi )=\Phi _{t}(y,\eta )\},
\end{equation*}
a general result, see e.g. Duistermaat \cite{DUI1} for a proof and
details, is :
\begin{theorem}
The operator $U_{h}(t)\Theta (P_{h})$ is an $h$-FIO associated to
$\Lambda$, there exists $U_{\Theta ,h}^{(N)}(t)$ with integral
kernel in $I(\mathbb{R}^{2n+1},\Lambda )$ and $R_{h}^{(N)}(t)$
bounded, with a $L^{2}$-norm uniformly bounded for $0<h\leq 1$ and
$t$ in a compact subset of $\mathbb{R}$, such that
$U_{h}(t)\Theta(P_{h})=U_{\Theta,h}^{(N)}(t)+h^{N}R_{h}^{(N)}(t)$.
\end{theorem}
The remainder, associated to $R_{h}^{(N)}(t)$, is under control
via the trick :
\begin{corollary}
If $\Theta _{1}\in C_{0}^{\infty }(I_{\varepsilon })$ is such that
$\Theta _{1}=1$ on $\rm{supp}(\Theta )$, then $\forall N\in
\mathbb{N}$ :
\begin{equation*}
\mathrm{Tr}(\Theta (P_{h})\varphi (\frac{P_{h}-E_{c}}{h}))=\frac{1}{2\pi }%
\mathrm{Tr}\int\limits_{\mathbb{R}}\hat{\varphi}(t)e^{\frac{i}{h}%
tE_{c}}U_{\Theta ,h}^{(N)}(t)\Theta
_{1}(P_{h})dt+\mathcal{O}(h^{N-n}).
\end{equation*}
\end{corollary}
The proof is easy by cyclicity of the trace (see e.g. \cite{Cam1} or \cite{[Rob]}).\medskip\\
In the particular case of Schr\"odinger operators it is well
known, via the BKW ansatz, that the integral kernel of
$U_{\Theta,h}^{(N)}$ can be written as :
\begin{equation}\label{BKW}
K_h^{(N)}(t,x,y)=\frac{1}{(2\pi h)^n}\int\limits_{\mathbb{R}^n}
b^{(N)}_h(t,x,y,\xi) e^{\frac{i}{h}(S(t,x,\xi)-\left\langle y,\xi
\right\rangle)} d\xi,
\end{equation}
where the function $S$ satisfies the  Hamilton-Jacobi equation :
\begin{equation*}
\partial _{t}S(t,x,\eta )+ p(x,\partial
_{x}S(t,x,\eta ))=0, \\
\end{equation*}
with initial condition $S(0,x,\xi)=\left\langle x,\xi
\right\rangle$. This imply that :
\begin{equation*}
\{(t,\partial _{t}S(t,x,\xi),x,\partial _{x}S(t,x,\xi),\partial
_{\xi}S(t,x,\xi),-\xi)\}\subset \Lambda ,
\end{equation*}
and that the function $S$ is a generating function of the flow,
i.e. :
\begin{equation}
\Phi _{t}(\partial _{\xi}S(t,x,\xi ),\xi ) =(x,\partial
_{x}S(t,x,\xi )). \label{Gene}
\end{equation}
Multiplying Eq. (\ref{BKW}) by $\hat{\varphi}(t)\exp(itE_c/h)$,
passing to trace and integrating w.r.t. the time yields :
\begin{equation}\label{gamma2 OIF}
\gamma_2(E_c,h,\varphi)\sim\frac{1}{(2\pi h)^n}\int\limits_{\mathbb{%
R\times R}^{2n}}e^{\frac{i}{h}(S(t,x,\xi )-\left\langle x,\xi
\right\rangle +tE_{c})}a^{(N)}_h(t,x,\xi
)\hat{\varphi}(t)dtdxd\xi,
\end{equation}
where $a^{(N)}_h(t,x,\xi)=\hat{\varphi}(t)b^{(N)}_h(t,x,x,\xi)$
and $b^{(N)}_h=b_0+hb_1+...+h^N b_N$, $\forall N$.
\begin{remark}
\rm{By a theorem of Helffer\&Robert, see e.g. \cite{[Rob]},
Theorem 3.11 and Remark 3.14, $\Theta (P_{h})$ is $h$-admissible
with a symbol supported in $p^{-1}(I_{\varepsilon })$. This allows
us to consider only oscillatory-integrals with compact
support.}$\hfill{\square}$
\end{remark}
\section{Classical dynamic near the equilibrium.}
Critical points of the phase function of (\ref{gamma2 OIF}) are
given by the equations :
\begin{equation*}
\left\{
\begin{array}{c}
\partial _{t}S(t,x,\xi )+E_{c}=0, \\
x=\partial _{\xi }S(t,x,\xi ), \\
\xi =\partial _{x}S(t,x,\xi ),
\end{array}
\right. \Leftrightarrow \left\{
\begin{array}{c}
p(x,\xi )=E_{c}, \\
\Phi _{t}(x,\xi )=(x,\xi ),
\end{array}
\right.
\end{equation*}
where the right hand side defines a closed trajectory of the flow
inside $\Sigma _{E_{c}}$. To obtain the contribution of the
critical point, we choose a function $\psi \in C_{0}^{\infty
}(T^{\ast }\mathbb{R}^{n})$, with $\psi =1$ near $z_{0}$ and such
that $\mathrm{supp}(\psi)\cap \Sigma_{E_c}=\{z_0\}$. Hence :
\begin{gather*}
\gamma _{2}(E_{c},h,\varphi) =\frac{1}{2\pi }\mathrm{Tr}\int\limits_{\mathbb{R}}e^{i%
\frac{tE_{c}}{h}}\hat{\varphi}(t)\psi ^{w}(x,hD_{x})\mathrm{exp}(-\frac{it}{h}%
P_{h})\Theta (P_{h})dt\\
+\frac{1}{2\pi }\mathrm{Tr}\int\limits_{\mathbb{R}}e^{i\frac{tE_{c}}{h}}\hat{%
\varphi}(t)(1-\psi
^{w}(x,hD_{x}))\mathrm{exp}(-\frac{it}{h}P_{h})\Theta (P_{h})dt,
\end{gather*}
where the asymptotics of the second term is given by the
semi-classical trace formula on a regular level. The first term,
which is precisely the quantity $\gamma_{z_0}(E_c,h,\varphi)$
studied in Theorem \ref{Main}, is micro-locally supported near
$z_0$ and the only contribution for this term, when $h$ tends to
0, arises from the set $\{(t,z_0),\text{
}t\in\mathrm{supp}(\hat{\varphi})\}$.

Now, we restrict our attention to $\gamma_{z_0}(E_c,h,\varphi)$.
Since $z_0$ is an equilibrium, by a standard argument in classical
mechanics, we obtain that :
\begin{equation}
d_{x,\xi}\Phi _{t}(z_{0})=\mathrm{exp}(tH_{-\Delta}),\text{
}\forall t.
\end{equation}
The computation of this linear operator is easy and gives :
\begin{equation}
d_{x,\xi}\Phi_{t}(z_{0})(u,v)=(u+2tv,v).
\end{equation}
A well known result is that the singularities w.r.t. $(x,\xi)$, in
the sense of the Morse lemma, of the function
$S(t,x,\xi)-\left\langle x,\xi\right\rangle$ are supported in the
set :
\begin{equation*}
\mathfrak{F}_t=\mathrm{Ker}(d_{x,\xi}\Phi_t(z_0)-\mathrm{Id}),
\end{equation*}
see e.g. Lemma 9 of \cite{Cam}. As mentioned in the introduction
we obtain :
\begin{gather*}
\mathfrak{F}_0 = T_{z_0} (T^* \mathbb{R}^{n})\simeq \mathbb{R}^{2n},\\
\mathfrak{F}_t = \{ (u,v) \in T_{z_0} (T^* \mathbb{R}^{n}) \text{
/ } v=0 \} \simeq \mathbb{R}^{n}, \text{ }t\neq 0.
\end{gather*}
To simplify notations, and until further notice, all derivatives
$d$ will be taken with respect to initial conditions $(x,\xi)$. We
can also identify each total derivatives $d^lf(z_0)$ with a form
of degree $l$ via :
\begin{equation*}
d^l f(z_0)(z-z_0)^l=\sum\limits_{|\alpha|=l} \partial^\alpha
f(z_0)(z-z_0)^{\alpha}.
\end{equation*}
The next terms of the germ of the flow in $z_0$ are computed via :
\begin{lemma}
\label{TheoFormule de récurence du flot}Let be $z_{0}$ an
equilibrium of the $C^{\infty}$ vector field $X$ and $\Phi _{t}$
the flow of $X$. Then for all $m\in \mathbb{N}^{\ast }$, there
exists a polynomial map $P_{m}$, vector valued and of degree at
most $m$, such that :
\begin{equation*}
d^{m}\Phi _{t}(z_{0})(z^{m})=d\Phi
_{t}(z_{0})\int\limits_{0}^{t}d\Phi _{-s}(z_{0})P_{m}(d\Phi
_{s}(z_{0})(z),...,d^{m-1}\Phi _{s}(z_{0})(z^{m-1}))ds.
\end{equation*}
\end{lemma}
For a proof we refer to \cite{Cam} or \cite{Cam1}. Here we use
that our vector field is :
\begin{equation*}
H_p= 2\xi \frac{\partial}{\partial x} - \partial_x
V(x)\frac{\partial}{\partial \xi},
\end{equation*}
and we identify the linearized flow in $z_0$ with a matrix
multiplication operator :
\begin{equation*}
d\Phi_t (z_0)=
\begin{pmatrix}
1 & 2t \\
0 & 1
\end{pmatrix}.
\end{equation*}
Clearly, with the hypothesis $(H_3)$ we obtain the polynomials :
\begin{gather*}
P_j=0,\text{ }\forall j\in\{2,..,k-2\},\\
P_{k-1}(Y_1,...,Y_{k-2})=
\begin{pmatrix}
0\\
-d^{k-1} \nabla_x V (x_0)(Y_1 ^{k-1})
\end{pmatrix}
\neq 0.
\end{gather*}
Where the notation $Y_1^{l}$ stands for $(Y_1,...,Y_1)$ :
$l$-times. Inserting the definition of $d\Phi_{-s}(z_0)$ and
integration from 0 to $t$ yields :
\begin{equation}\label{derivative-flow}
d^{k-1} \Phi_t(z_0)((x,\xi)^{k-1})=\begin{pmatrix}
1 & 2t \\
0 & 1
\end{pmatrix}
\int\limits_{0}^{t}
\begin{pmatrix}
2s d^{k-1} \nabla_x V (x_0)((x+2s\xi)^{k-1})\\
-d^{k-1} \nabla_x V (x_0)((x+2s\xi)^{k-1})
\end{pmatrix}
ds.
\end{equation}
If $z_0=0$ the germ of order $k-1$ of the flow can be written as :
\begin{equation}\label{germ-flot}
\Phi_t(z)=d\Phi_t(0)(z)+\frac{1}{(k-1)!}
d^{k-1}\Phi_t(0)(z^{k-1})+\mathcal{O}(||z||^k),
\end{equation}
and can be computed explicitly. Terms of higher degree can be
recursively obtained but we do not need them for the present work.
\section{Normal forms of the phase function.}
Since the contribution we study is micro-local, in a neighborhood
$V(z_0)$ of the critical point we can work in local coordinates
and we identify $T^{\ast }\mathbb{R}^{n}\cap V(z_0)$ with an open
of $\mathbb{R}^{2n}$. We define :
\begin{equation}
\Psi(t,z)=\Psi(t,x,\xi)=S(t,x,\xi )-\left\langle x,\xi
\right\rangle+tE_{c},\text{ }z=(x,\xi)\in \mathbb{R}^{2n}.
\label{defphase}
\end{equation}
We start by a more precise description of our phase function.
\begin{lemma}\label{structure phase} Near $z_0$, here supposed to be 0 to simplify, we
have :
\begin{equation}
\Psi(t,x,\xi)=-t||\xi||^2+ S_k(t,x,\xi)+R_{k+1}(t,x,\xi ),
\label{forme phase}
\end{equation}
where $S_k$ is homogeneous of degree $k$ w.r.t. $(x,\xi)$ and is
uniquely determined by $V_k$. Moreover
$R_{k+1}(t,x,\xi)=\mathcal{O}(||(x,\xi )||^{k+1})$ uniformly for
$t$ in a compact.
\end{lemma}
\textit{Proof.} In view of Eq. (\ref{germ-flot}), it is natural to
try :
\begin{equation*}
S(t,x,\xi)=-tE_c+S_2(t,x,\xi)+S_k(t,x,\xi)+\mathcal{O}(||(x,\xi)||^{k+1}),
\end{equation*}
where the $S_j$ are time dependant and homogeneous of degree $j$
w.r.t. $(x,\xi)$. Combining
$\Phi_t(\partial_{\xi}S(t,x,\xi),\xi)=(x,\partial_{x}S(t,x,\xi))$
and Eq. (\ref{germ-flot}), we obtain first :
\begin{equation*}
S_2(t,x,\xi)=\left\langle x,\xi \right\rangle-t||\xi ||^2.
\end{equation*}
Similarly, if we retain only terms homogeneous of degree $k-1$, we
get :
\begin{equation*}
d\Phi_t(0)((\partial_\xi S_k,0))+\frac{1}{(k-1)!}d^{k-1}
\Phi_t(0)((\partial_\xi S_2,\xi)^{k-1})=(0,\partial_x S_k).
\end{equation*}
If $J$ is the matrix of the usual symplectic form $\sigma$ we
rewrite this as :
\begin{equation*}
J\nabla S_k(t,x,\xi)=\frac{1}{(k-1)!}d^{k-1}\Phi
_{t}(0)((x-2t\xi,\xi)^{k-1}).
\end{equation*}
By homogeneity we obtain :
\begin{equation*}
S_k(t,x,\xi)=\frac{1}{k!}\sigma ((x,\xi),d^{k-1} \Phi_t(0)
((x-2t\xi,\xi)^{k-1})),
\end{equation*}
and this gives the result since $d^{k-1} \Phi_t(0)$ is given by
Eq. (\ref{derivative-flow}). $\hfill{\blacksquare}$\medskip\\
Fortunately, we will not have to compute the remainder explicitly
because of some homogenous considerations. Now, we study carefully
the function $S_k$.
\begin{corollary} \label{structure fine}
The function $S_k$ satisfies :
\begin{equation}
S_k(t,x,\xi)=-t V_k(x) +t^2 \left\langle \xi, \nabla_x
V_k(x)\right\rangle +\sum\limits_{j,l=1}^{n} \xi_j\xi_l
g_{j,l}(t,x,\xi),
\end{equation}
where the functions $g_{jl}$ are smooth and vanish in $(x,\xi)=0$.
\end{corollary}
\textit{Proof.} If $f$ is homogeneous of degree $k>2$ w.r.t.
$(x,\xi)$ we can write :
\begin{equation*}
f(x,\xi)=f_1(x)+\left\langle \xi, f_2(x) \right\rangle +
\sum\limits_{j,l=1}^{n} \xi_j\xi_l f_3^{(jl)}(x,\xi).
\end{equation*}
A similar result is valid if $f$ has time dependant coefficients.
Hence, it remains to compute terms of degree 0 and 1 w.r.t. $\xi$.
By construction we have :
\begin{gather*}
S_k(t,x,\xi)=\frac{1}{k!}
\sigma((x,\xi),d\Phi_t(0)\int\limits_{0}^{t}
\begin{pmatrix}
2s d^{k-1} \nabla_x V(0)((x+2(s-t)\xi)^{k-1})\\
-d^{k-1} \nabla_x V(0)((x+2(s-t)\xi)^{k-1})
\end{pmatrix}
ds.
\end{gather*}
Clearly, the term homogeneous of degree $k$ w.r.t. $x$ is given by
:
\begin{equation}
-\frac{1}{k!}\int\limits_{0}^t \left\langle x, d^{k-1} \nabla_x
V(0)(x^{k-1})\right\rangle ds=-\frac{t}{k} \left\langle x,
\nabla_xV_k(x) \right\rangle =-tV_k(x).
\end{equation}
By combinatoric and linear operations, the term linear in $\xi$
can be written :
\begin{gather*}
\frac{2}{k!}\left( (k-1) \left\langle x,d^{k-1} \nabla_x
V(0)((x^{k-2},\xi))\right\rangle +\left\langle \xi, d^{k-1}
\nabla_x V(0)(x^{k-1}) \right\rangle\right )\int\limits_{0}^t
(s-t)ds\\
=\frac{t^2}{k}\left( (k-1) \left\langle \nabla_x V_k(x),\xi
\right\rangle +\left\langle \xi,\nabla_x V_k(x)
\right\rangle\right )=t^2 \left\langle \xi, \nabla_x V_k (x)
\right\rangle.
\end{gather*}
This complete the proof. $\hfill{\blacksquare}$\medskip\\
These two terms can be also derived more intuitively. From the
Hamilton-Jacobi equation we obtain $S(0,x,\xi)=\left\langle
x,\xi\right\rangle$, $\partial_tS(0,x,\xi)=-p(x,\xi)$  and :
\begin{equation*}
\partial^2_{t,t} S(t,x,\xi)=-\left\langle \partial_\xi p(x,\partial_x S(t,x,\xi)),
\partial^2_{t,x}S(t,x,\xi)\right\rangle.
\end{equation*}
We get $\partial^2_{t,t} S(0,x,\xi)= 2\left\langle \xi,
\partial_x V(x)\right\rangle$ but this method gives no clear
information about the degree w.r.t. $(x,\xi)$ of the remainders
$\mathcal{O}(t^d)$, $d\geq 3$.

We have now enough material to establish the normal of our phase
function.
\begin{lemma}\label{FN1}
In a neighborhood of $(t,z)=(t,z_0)$, there exists local
coordinates $y$ such that :
\begin{equation}
\Psi(t,z) \simeq -y_{0}(y_1^2+y_{2}^{k}).
\end{equation}
\end{lemma}
\noindent\textit{Proof.} We can here assume that $z_{0}$ is the
origin. We proceed in two steps. First we want to eliminate terms
of high degree. Let be :
\begin{equation*}
E(t,x,\xi)=\sum\limits_{j,l}\xi_j \xi_l g_{j,l}(t,x,\xi).
\end{equation*}
Since $S(t,x,\xi)-\left\langle x,\xi\right\rangle=\mathcal{O}(t)$,
we have $\Psi(t,z)=\mathcal{O}(t)$. Hence, all terms of the
expansion are $\mathcal{O}(t)$ and this allows to write
$E=t\tilde{E}$ and $R_{k+1}=t\tilde{R}_{k+1}$. We use polar
coordinates $x=r\theta$, $\xi=q\eta$, $\theta,\eta \in
\mathbb{S}^{n-1}(\mathbb{R})$, $q,r\in\mathbb{R}_{+}$. This
induces naturally a jacobian $r^{n-1}q^{n-1}$. By construction,
there exists a function $F$ vanishing in $(q,r)=(0,0)$ such that :
\begin{equation}
\tilde{E}(t,r\theta,q\eta)=q^2 F(t,r,\theta,q,\eta).
\end{equation}
With Lemma \ref{structure phase} and Corollary \ref{structure
fine}, near the critical point we have :
\begin{equation*}
\Psi(t,z) \simeq -t\left(q^{2} +r^k V_k(\theta)-t
qr^{k-1}\left\langle \eta,\nabla V_k(\theta) \right\rangle +q^2
F(t,r,\theta,q,\eta)+\tilde{R}_{k+1}(t,r\theta,q\eta)\right ).
\end{equation*}
Thanks to the Taylor formula, the remainder $\tilde{R}_{k+1}$ can
be written as :
\begin{equation*}
\tilde{R}_{k+1}(t,r\theta,q\eta)=q^2
R_1(t,r,\theta,q,\eta)+r^{k}R_2(t,r,\theta,q,\eta),
\end{equation*}
where $R_1$ vanishes in $r=0$ and $R_2$ vanishes in $q=0$. We
obtain :
\begin{equation*}
\Psi(t,z) \simeq -t(q^{2}\alpha_1(t,r,\theta,q,\eta) +r^k
\alpha_2(t,r,\theta,q,\eta))+t^2 qr^{k-1}\left\langle \eta,\nabla
V_k(\theta) \right\rangle,
\end{equation*}
with :
\begin{gather*}
\alpha_1(t,r,\theta,q,\eta)=(1+R_1+\tilde{F})(t,r,\theta,q,\eta),\\
\alpha_2(t,r,\theta,q,\eta)=V_k(\theta)+R_2(t,r,\theta,q,\eta).
\end{gather*}
Since $V_k(\theta)>0$, we can eliminate $\alpha_1$ and $\alpha_2$
by a local change of coordinates :
\begin{equation} \label{alpha}
(q \alpha_1^{\frac{1}{2}}, r \alpha_2^{\frac{1}{k}})\rightarrow
(q,r),
\end{equation}
near $(q,r)=(0,0)$. We obtain the new phase :
\begin{equation}
\Psi(t,z)\simeq -t(q^{2} +r^k)+t^2
qr^{k-1}\varepsilon(t,r,\theta,q,\eta),
\end{equation}
where :
\begin{equation*}
\varepsilon (t,r,\theta,q,\eta)= \left\langle \eta,\nabla
V_k(\theta)
 \right\rangle(\alpha_1^{-\frac{1}{2}}
\alpha_2^{\frac{1-k}{k}})(t,r,\theta,q,\eta).
\end{equation*}
In a second time, we eliminate the nonlinear term in $t$. To do
so, we write :
\begin{equation*}
-t(q^2-tqr^{k-1}\varepsilon)=-t(q-\frac{t}{2}r^{k-1}\varepsilon)^2
+\frac{1}{4}t^3 r^{2k-2}\varepsilon^2.
\end{equation*}
Finally, we can factor out the last term. To do so, we define :
\begin{equation*}
\alpha_3(t,r,\theta,q,\eta)=(1-\frac{t^2}{4}r^{k-2}\varepsilon^2(t,r,\theta,q,\eta)).
\end{equation*}
Then the change of variables $r\rightarrow
\alpha_3^{-\frac{1}{k}}r$ gives :
\begin{equation*}
\Psi(t,z)\simeq
-t((q-\frac{t}{2}r^{k-1}\tilde{\varepsilon}(t,r,\theta,q,\eta))^2
+r^k),
\end{equation*}
where :
\begin{equation*}
\tilde{\varepsilon}(t,r,\theta,q,\eta)=
\alpha_3^{\frac{1-k}{k}}\varepsilon(t,\alpha_3^{-\frac{1}{k}}r,t,\theta,q,\eta).
\end{equation*}
Finally, we define :
\begin{gather}(y_0,y_2,y_3,...,y_{2n})(t,r,\theta,q,\eta) =(t,r,\theta,\eta),\\
y_1(t,r,\theta,q,\eta)=
q-\frac{t}{2}r^{k-1}\tilde{\varepsilon}(t,r,\theta,q,\eta),
\end{gather}
to get the desired result. $\hfill{\blacksquare}$\medskip\\
If we use these local coordinates we obtain a simpler problem :
\begin{equation}\label{simpleIO}
\int\limits_{\mathbb{R}\times\mathbb{R}_{+}^2}
e^{-\frac{i}{h}y_0(y_1^2+y_2^k)}A(y_0,y_1,y_2) dy_0dy_1 dy_2,
\end{equation}
where this new amplitude $A$ is obtained via :
\begin{equation}
A(y_0,y_1,y_2)=\int y^{*}(a |Jy|) dy_3...dy_{2n}.
\end{equation}
\begin{remark}\label{Jacobian}
\rm{Eq. (\ref{simpleIO}) describes correctly the problem since the
critical set is :
\begin{equation}
\mathfrak{C}(-y_0(y_1^2+y_2^k))=\{ (y_0,0,0),\text{ } y_0\in
\mathbb{R}\},
\end{equation}
and the new phase is non-degenerate w.r.t. $y_1$ when $y_0\neq 0$.
To attain our objective, we must see that we have
$(y_1,y_2)(t,r,\theta,q,\eta)=(0,0) \Leftrightarrow (x,\xi)=z_0$
and that the introduction of the polar coordinates yields :
\begin{equation*}
A(y_0,y_1,y_2)=\mathcal{O}(y_l ^{n-1}),\text{ }l=1,2.
\end{equation*}
Finally, all the diffeomorphism used have Jacobian 1 in $z_0$
excepted the correction w.r.t. $\alpha_2$, cf. Eq. (\ref{alpha}),
which induces a multiplication by :
\begin{equation}
|V_k(\theta)|^{-\frac{n}{k}}.
\end{equation}
These facts will be implicitly used in the next
section.}$\hfill{\square}$
\end{remark}
\section{Proof of the main result.}
Let be $k$ the even integer attached to our potential. We start by
a lemma which allows to compute the asymptotic expansion of our
oscillatory integrals.
\begin{lemma}\label{DA-IO}
For $a\in C_0^\infty(\mathbb{R}\times [0,\infty]^2)$, the
asymptotic equivalent :
\begin{equation}
\int\limits_{\mathbb{R}\times\mathbb{R}_{+}\times \mathbb{R}_{+}}
e^{-i\lambda y_0 (y_1^2+y_2^k)}a(y_0,y_1,y_2)dy_0 dy_1 dy_2\sim
\sum\limits_{j,l\geq 0} \lambda^{-\frac{j+1}{2}-\frac{l+1}{k}}
C_{j,l}(a),
\end{equation}
holds, as $\lambda\rightarrow +\infty$, with the distributional
coefficients :
\begin{equation}
C_{j,l}(a)=  \int \limits_{\mathbb{R}_{+}\times \mathbb{R}_{+}}
\frac{\partial^j
\partial^l \hat{a}}{\partial y_1^j \partial y_2^l} (y_1^2+y_2^k,0,0) y_1^j y_2^l
dy_1dy_2,
\end{equation}
where $\hat{a}$ is the partial Fourier transform of $a$ w.r.t.
$y_0$.
\end{lemma}
\textit{Proof.} Let us consider first an amplitude
$a_1(y_0)a_2(y_1) a_3(y_2)$. We note $I(\lambda)$ our integral,
changing $(y_1,y_2)$ by
$(\lambda^{-\frac{1}{2}}y_1,\lambda^{-\frac{1}{k}}y_2)$ and
integrating w.r.t. $y_0$ yields :
\begin{equation*}
I(\lambda)= \lambda^{-\frac{1}{2}-\frac{1}{k}}\int
\limits_{\mathbb{R}_{+}\times \mathbb{R}_{+}}
\hat{a_1}(y_1^2+y_2^k)a_2(\frac{y_1}{\lambda^{\frac{1}{2}}})
a_3(\frac{y_2}{\lambda^{\frac{1}{k}}})dy_1dy_2.
\end{equation*}
Since  $\hat{a_1}(y_1^2+y_2^k)$ decreases faster than any
polynomial when $y_1$ or $y_2$ tends to infinity we insert finite
Taylor expansions for $a_2$ and $a_3$ to obtain :
\begin{equation*}
I(\lambda)= \sum\limits_{j=0}^J \sum\limits_{l=0}^L
\lambda^{-\frac{j+1}{2}-\frac{l+1}{k}} \int
\limits_{\mathbb{R}_{+}\times \mathbb{R}_{+}}
\hat{a_1}(y_1^2+y_2^k) \frac{a_2^{(j)}(0)}{j!}
\frac{a_3^{(l)}(0)}{l!} y_1^j y_2^l dy_1 dy_2+R_{J,L}(\lambda).
\end{equation*}
Since $k>2$, easy estimates and support considerations show that :
\begin{equation*}
R_{J,L}(\lambda)=\mathcal{O}(\lambda^{-\frac{J+1}{2}-\frac{L+2}{k}}).
\end{equation*}
Finally, since the coefficients obtained are continuous linear
functionals, the general formula holds for $a\in
C_0^\infty(\mathbb{R}\times [0,\infty]^2)$. $\hfill{\blacksquare}$
\begin{remark} \rm{This lemma holds also for
integration w.r.t $(y_1,y_2)$ on $\mathbb{R}^2$. But this lemma
cannot be applied for a difference $y_1^2 -y_2^k$ since the term
$\hat{\varphi}(y_1^2 -y_2^k)$ is equal to $\hat{\varphi}(0)$ on
the set $y_1 =\pm y_2^{\frac{k}{2}}$.}$\hfill{\square}$
\end{remark}
We apply Lemma \ref{DA-IO} to our amplitude $A(y_0,y_1,y_2)$ to
prove the existence of the complete asymptotic expansion. Using
Remark \ref{Jacobian}, we obtain that the first non-zero
coefficient of this expansion occurs for $j=l=n-1$ with :
\begin{equation*}
C_{n-1,n-1}(A)=\frac{1}{((n-1)!)^2}\int
\limits_{\mathbb{R}_{+}\times \mathbb{R}_{+}} \frac{\partial^{n-1}
\partial^{n-1} \hat{A}}{\partial y_1^{n-1} \partial y_2^{n-1}} (y_1^2+y_2^k,0,0) (y_1 y_2)^{n-1}
d y_1dy_2,
\end{equation*}
and the corresponding contribution is :
\begin{equation}
I(\lambda)\sim \lambda ^{-\frac{n}{2}-\frac{n}{k}}C_{n-1,n-1}(A)
+\mathcal{O}(\lambda^{-\frac{n}{2}-\frac{n+1}{k}}).
\end{equation}
Now we must express this amplitude with the data of the problem.
To avoid unnecessary calculations, we introduce the new amplitude
:
\begin{equation*}
A(y_0,y_1,y_2)=(y_1 y_2)^{n-1} \tilde{A}(y_0,y_1,y_2).
\end{equation*}
By substitution and inversion of the Fourier transform we get :
\begin{gather}
\frac{2\pi}{((n-1)!)^2}\int \limits_{\mathbb{R}_{+}\times
\mathbb{R}_{+}} \frac{\partial^{n-1}
\partial^{n-1} \hat{A}}{\partial y_1^{n-1} \partial y_2^{n-1}} (u^2+v^k,0,0) (uv)^{n-1}
du dv \notag \\
=\int\limits_{\mathbb{R}\times\mathbb{R}_{+}^2} e^{-it(u^2+v^k)}
\tilde{A}(t,0,0) (uv)^{n-1} dtdudv. \label{IO2}
\end{gather}
But by construction we have :
\begin{equation*}
\tilde{A}(y_0,y_1,y_2)=\int y^* (a|Jy|(t,r\theta,q\eta)) (y_1
y_2)^{1-n} dy_3...dy_{2n}.
\end{equation*}
Easy manipulations and the properties of the coordinates $y$ (cf.
Lemma \ref{FN1} and Remark \ref{Jacobian}) show that :
\begin{equation}\label{ampli}
\tilde{A}(y_0,0,0)=\hat{\varphi}(y_0)\int\limits_{\mathbb{S}^{n-1}\times
\mathbb{S}^{n-1}} |V_k(\theta)|^{-\frac{n}{k}} d\eta d\theta.
\end{equation}
Setting $\lambda=h^{-1}$, dividing by $(2\pi h)^n$  and
substituting Eq. (\ref{ampli}) in Eq. (\ref{IO2}) we obtain the
result stated in Theorem \ref{Main}.
$\hfill{\blacksquare}$\medskip\\
\textbf{Extension and example.}\\
The main result can be extended
to the case of an $h$-admissible operator. If $p_1$ is the
sub-principal symbol the half-density of the propagator is given
by :
\begin{equation}
\exp(i\int\limits_{0}^t p_1(\Phi_s(y,-\eta))ds)
|dtdyd\eta|^{\frac{1}{2}},
\end{equation}
see \cite{D-H} concerning the solution of the first transport
equation. But the critical point is invariant under the flow and
if $p_1$ vanishes at the critical point the statement is the same.
Otherwise, the first term of the BKW expansion satisfies
$a^{(0)}(t,z_0)=\hat{\varphi}(t) \exp(itp_1(z_0))$ and this causes
a shift by $p_1(z_0)$ in Theorem \ref{Main}.\\
The Witten Laplacian on 0-forms is given by :
\begin{equation}
\Delta^{(0)}_{h,f}=-h^2 \Delta +|\nabla f(x)|^2 -h\Delta f(x),
\end{equation}
we refer to \cite{Hel} for the signification and applications in
statistical mechanics of this operator. This is an
$h$-pseudodifferential operator with $p_0(x,\xi)=\xi^2 +|\nabla
f(x)|^2$ and $p_1(x,\xi)=-\Delta f(x)$. The fixed points of the
classical system, attached to $p_0$, are given by the critical
points of $f$ with 0 momentum. When these are not degenerate the
associated trace formula can be derived by using the results of
\cite{BPU} for the contribution of the $0$ period and for strictly
positive periods of the linearized flow from \cite{Cam} or
\cite{KhD1}. Note that the corresponding quadratic approximation
is always elliptic in this setting.\medskip\\
An interesting problem, see \cite{Hel} p.33, is the transition
from the double well to a single degenerate well. For example one
can consider the family of functions :
\begin{equation}
f_s(x)=(s-1)||x||^2 +||x||^4, \text{ }s\in [0,1],\text{ }
x\in\mathbb{R}^n.
\end{equation}
For the critical value $s=1$ we obtain that $|\nabla
f_1(x)|^2=16||x||^6$ and since $p_1(0,\xi)=-\Delta f_1(0)=0$ we
can apply Theorem \ref{Main}, here in an exact version, to compute
the trace formula for the critical energy $E_c=0$ :
\begin{equation*}
\gamma(0,h,\varphi)=h^{-n+\frac{2n}{3}}(\frac{\mathrm{S}(\mathbb{S}^{n-1})^2}{(2\pi)^n}2^{-\frac{4n}{6}}
\int\limits_{\mathbb{R}_{+}\times \mathbb{R}_{+}}\varphi
(u^2+q^6)u^{n-1} v^{n-1}dudv+\mathcal{O}(h^{\frac{1}{6}})).
\end{equation*}
The transition probability, for the associated measure $\exp
(-f_s(x)/h)dx$, is established in
\cite{Hel} via Laplace transform considerations.\bigskip\\
\textbf{Acknowledgements.} This work was partially supported by
the \textit{IHP-Network}, reference number HPRN-CT-2002-00277 and
by the \textit{SFB/TR12} project, \textit{Symmetries and
Universality in Mesoscopic Systems}.

\end{document}